# Importance of Intent-Sharing for V2X-based Maneuver Coordination

Rafael Molina-Masegosa[a], Sergei S. Avedisov[b], Miguel Sepulcre[a],
Javier Gozalvez[a], Yashar Z. Farid[b1], Onur Altintas[b]
[a] Universidad Miguel Hernandez de Elche, Elche, Spain, {rafael.molinam, msepulcre, j.gozalvez}@umh.es
[b] Toyota Motor North America R&D – InfoTech Labs, Mountain View, CA, USA, {sergei.avedisov, yashar.zeyinali.farid, onur.altintas}@toyota.com

***Abstract*— This paper examines the critical role of intent-sharing in enabling effective maneuver coordination for connected and automated vehicles (CAVs). Successful maneuver coordinations require vehicles to accurately know other vehicles' driving intentions. Intent-sharing can be achieved by the remote vehicles directly communicating their plans with the ego vehicle, as opposed to the ego vehicle predicting the trajectory on the remote vehicles' behalf. In this paper, we investigate the potential of intent-sharing on maneuver coordination effectiveness by quantifying the percentage of successful coordinations. We analyze the potential of intent-sharing by comparing its effectiveness for coordinated lane changes in a highway scenario with the effectiveness of a trajectory prediction method based on current kinematic data. Our analysis demonstrates in two scenarios substantial improvements in maneuver coordination when CAVs have direct access to the nearby vehicles' driving intentions through intent sharing. These findings highlight the importance of including intent-sharing in the maneuver coordination protocol.**

## I. Introduction

The coordination of driving maneuvers (a.k.a. maneuver coordination or cooperative maneuvering) is a cornerstone of connected and automated driving. By leveraging vehicle-to-everything (V2X) communications, vehicles can collaborate to perform maneuvers such as lane changes, merges, and overtakes. However, the effectiveness of these maneuvers heavily depends on the ability of vehicles to estimate future actions of nearby vehicles. For example, in cooperative driving, vehicles can predict the trajectory of nearby vehicles using information about their position and speed that is included in exchanged CAMs (Cooperative Awareness Messages), BSMs (Basic Safety Messages). CPMs (Collective Perception Messages) and SDSMs (Sensor Data Sharing Messages) may serve as additional sources of data for trajectory prediction. However, trajectory predictions are prone to estimation errors due to the need to infer other vehicles' intentions and the complexity and variability of traffic scenarios, which can lead to maneuver coordination failures.

Intent-sharing [1] addresses this limitation by allowing vehicles to directly communicate their planned trajectories, providing a more accurate basis for decision-making. Intent-sharing has been standardized in SAE [1] as optional in MSCMs (Maneuver Sharing and Coordination Messages). In ETSI, intent-sharing is under discussion in [2] for maneuver coordination in MCM (Maneuver Coordination Messages). In addition, ETSI recently decided to optionally include in the Release 2 of the CAM the future trajectory of the transmitting vehicle (once every second approximately) [3], which is in line with the intent-sharing concept. Also, [4] demonstrated with experimental results that intent-sharing can reduce future trajectory uncertainty.

In this study, we analyze the potential of intent-sharing to improve maneuver coordinations. The main motivation is the lack of large-scale quantitative studies analyzing the benefits of intent-sharing in predicting and enhancing maneuver coordination success. To this aim, we compare the execution of maneuver coordinations when vehicles predict the future trajectories of surrounding vehicles using their current kinematics (position and velocity), and when using intent-sharing. This analysis is performed under two distinct scenarios to demonstrate the potential of intent-sharing under different conditions. The accuracy of the trajectory prediction depends on the knowledge of the surrounding traffic, and the execution of maneuver coordinations can be impacted by prediction errors. With intent-sharing, vehicles broadcast their driving intentions (i.e. their planned trajectories) and vehicles do not need to predict the trajectory of surrounding vehicles to coordinate maneuvers. In this study, we demonstrate that CAVs can better coordinate their driving maneuvers by knowing other vehicles' driving intentions thanks to intent-sharing, reducing uncertainty and improving the overall success rate of coordinated maneuvers.

The rest of this paper is organized as follows. Section II explains why intent-sharing can improve the effectiveness of maneuver coordination. Section III details the methodology used in this work to evaluate the impact of intent-sharing on maneuver coordination, including the simulation framework and the maneuver coordination implementation. Section IV quantifies the improvements in maneuver coordinations achieved with intent-sharing compared to a baseline trajectory prediction.

## II. The Role of Intent Sharing in Maneuver Coordination

Effective maneuver coordination relies heavily on accurately knowing the near-future movements (within a few seconds) of surrounding vehicles. Before triggering the maneuver coordination process, a vehicle must assess its

This work has been partially funded by MICIU/AEI/10.13039/501100011033 and "ERFD/EU" (PID2023-150308OB-I00), and the "European Union NextGenerationEU/PRTR" (TED2021-130436B-I00), and UMH's Vicerrectorado de Investigación grants.

[1] This research was conducted while Yashar Z. Farid was at Toyota Motor North America R&D – InfoTech Labs.

.

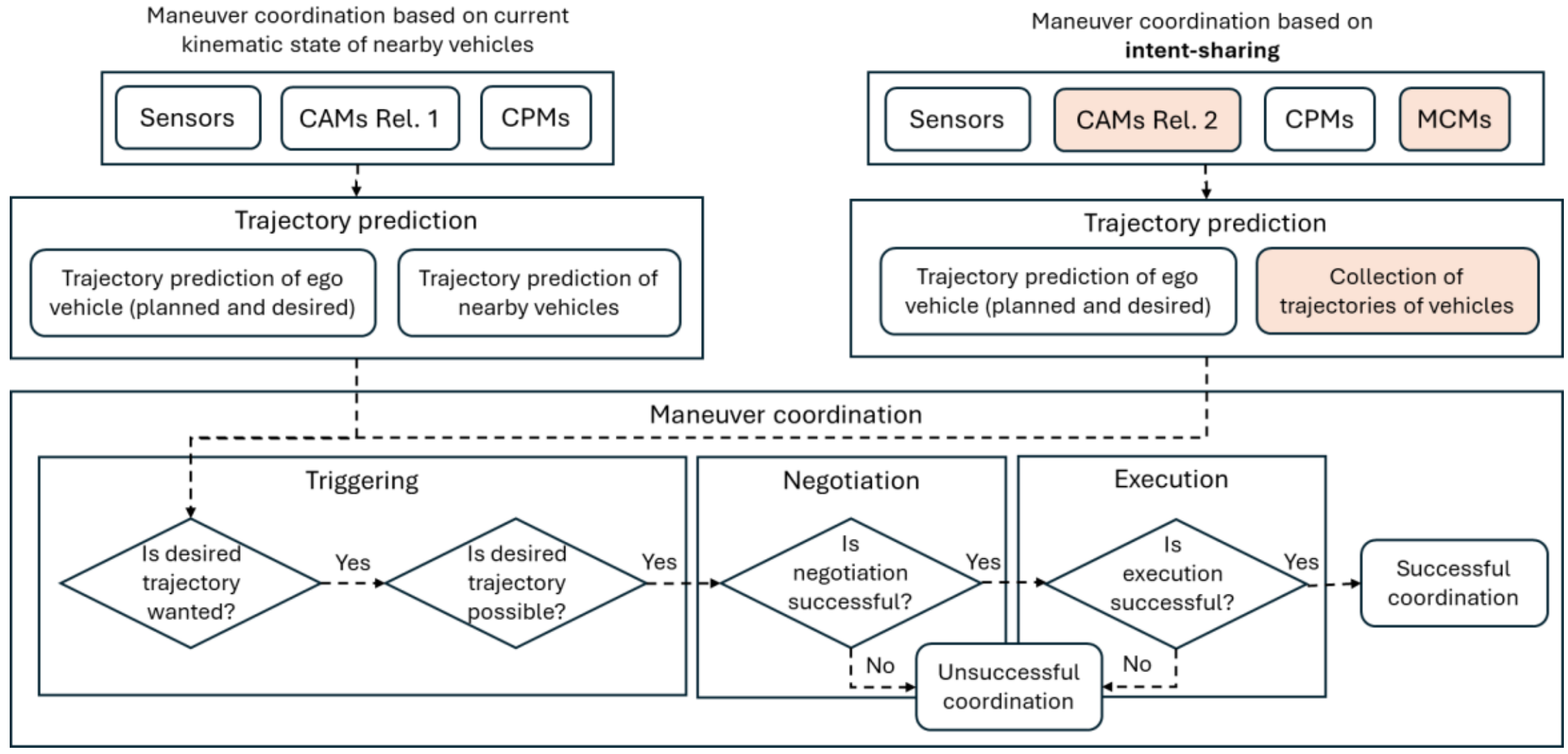

Fig. 1. Maneuver coordination with and without intent sharing. Red shading highlights elements introduced by intent-sharing. Naming conventions follow ETSI standards.

current planned trajectory, i.e. the path it intends to follow without requiring cooperation from other vehicles [5]. Furthermore, it should identify potentially more advantageous alternative trajectories, i.e. desired trajectories [5], and determine their feasibility. When the desired trajectory cannot be followed, e.g. due to the presence of an obstructing vehicle, a maneuver coordination can be triggered. The decision to trigger a maneuver coordination therefore depends on the predicted trajectories of both the ego vehicle and the surrounding vehicles. As a consequence, the accuracy of trajectory prediction could significantly impact the successful execution of coordinated maneuvers. In addition, it could influence the ability of the ego vehicle to assess the potential benefits of a maneuver coordination, which prevents unnecessary coordinations that may actually not benefit the ego vehicle.

Various trajectory prediction methods exist in the literature, ranging from physics-based models to machine learning approaches [6]. Regardless of the specific method, prediction accuracy improves with greater knowledge of the surrounding traffic [6]. The knowledge about the position and speed of surrounding vehicles could be obtained from onboard sensors or V2X data such as CAM/BSMs as well as CPMs (Collective Perception Messages). However, with intent sharing, vehicles do not need to predict the trajectory of surrounding vehicles, as the surrounding vehicles will directly broadcast their planned trajectory, and hence provide the most precise data about their current and near-future states. Fig. 1 illustrates how intent-sharing can be used as input for trajectory prediction, and consequently, maneuver coordination triggering. Each vehicle uses trajectory prediction to generate its own planned trajectory. This planned trajectory is then regularly broadcast to all surrounding vehicles through MCMs or MSCMs (or CAMs in ETSI Release 2 [3]). All vehicles leverage this shared information for their own trajectory predictions, resulting in more accurate planned trajectories. The exchanged MCMs are then used to improve the trajectory prediction, which in turn will be included in the next MCM to improve other vehicles' trajectory prediction. This feedback loop has the potential to significantly improve trajectory prediction, which is critical for effective maneuver coordination, as previously discussed.

## III. Methodology

We analyze the potential of intent sharing to improve maneuver coordinations considering the case of lane change coordination on highways. We evaluate this potential by comparing two trajectory prediction approaches: a baseline prediction approach using constant speed and an intent-sharing based prediction approach.

### *A. Maneuver coordination triggering and trajectory prediction*

The decision to trigger a maneuver coordination requires the ego vehicle to continuously monitor its surrounding traffic. If certain conditions are met, the ego vehicle initiates a maneuver coordination by requesting cooperation from other vehicles. For the lane change maneuver implemented in this work (detailed in [7]), the following two conditions must be satisfied to trigger a coordination:

- The ego vehicle ($V_{ego}$) predicts that a lane change would be beneficial for itself within the next 5 seconds. However, the lane change is currently obstructed by another vehicle in the adjacent lane ($V_{adj}$), given the current planned trajectory of $V_{adj}$.
- $V_{ego}$ could safely and efficiently execute the desired lane change if $V_{adj}$ decelerates at a controlled rate of -2 m/s² during 1 second, followed by maintaining a constant speed for the remainder of the maneuver. This specific deceleration profile reflects the actual coordination execution behavior of $V_{adj}$. Therefore, this condition serves to verify in advance whether the lane change is feasible given the expected motion of $V_{adj}$.

While these triggering conditions are accurately defined, the process of verifying these rules relies heavily on precise trajectory prediction, as discussed in Section II. Accurate predictions enable the coordinated lane change maneuver to proceed as planned. Conversely, if the traffic evolves differently than initially estimated at the triggering time, the maneuver will likely be aborted, preventing the intended lane change. We consider two distinct prediction approaches for the maneuver coordination: baseline approach and intent-sharing based prediction approach.

The baseline prediction approach makes use of a physics-model based on kinematics with constant speed [6]. A model based on constant acceleration has also been evaluated with similar results. The model predicts the trajectories of only the nearest vehicles, in addition to the ego vehicle's own trajectory. These nearest vehicles include vehicles immediately adjacent to the ego vehicle in both its current lane and the target lane. This is the smallest set of vehicles whose predicted trajectories are required to evaluate the maneuver coordination triggering conditions. Trajectories are predicted under the assumption that these vehicles will maintain their current speeds. Their current kinematic states (position and speed) can be obtained through onboard sensors or CAMs.

The intent-sharing based prediction approach models an intent-sharing scenario where vehicles can accurately know their own trajectory and those of nearby vehicles. Knowing the exact trajectories of all vehicles in the scenario represents an upper bound scenario but of too high computational complexity. To reduce this complexity, our intent-sharing approach considers only a subset of the vehicles around the ego vehicle. In particular, it considers the ego vehicle and the vehicles immediately behind, as well as all vehicles ahead of the ego vehicle up to a distance $d_{pred}$ that is set to 500m (see Fig. 2). For all considered vehicles, trajectories are computed using the same mobility models employed by the traffic simulation platform for modeling the vehicles' mobility, including lane changes and accelerations/decelerations, except for the edge vehicles close to $d_{pred}$ for which a constant speed is assumed. The computed trajectories can have certain errors due to the limit imposed by $d_{pred}$ to achieve the necessary balance between accuracy and computational complexity. Other values for $d_{pred}$ have been evaluated from 400m to 600m, with similar trends and conclusions.

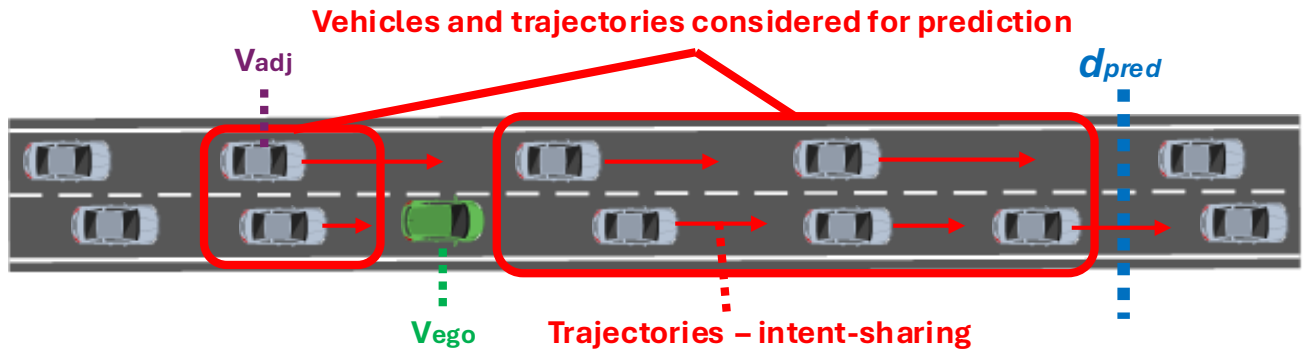

Fig. 2. Vehicles considered for trajectory prediction in the intent-sharing based prediction approach

### B. Maneuver coordination design

We implemented a maneuver coordination based on the design detailed in [8], adhering to the SAE Maneuver Sharing and Coordination Service standard [1]. This implementation uses a state machine model and a structured protocol involving a Host Vehicle (HV) and a Remote Vehicle (RV). For simplicity, each HV coordinates with one RV, In our lane change scenario, $V_{ego}$ acts as the HV, and $V_{adj}$ is the RV (Fig. 2). The process involves distinct states (awareness, negotiation and execution) and message exchanges. Upon satisfying the maneuver coordination triggering conditions (as described in Section III.A), the HV ($V_{ego}$) enters the negotiation state. During negotiation, the HV repeatedly broadcasts *Request* messages (every 100ms) to the RV, signaling its desire to perform a specific coordination maneuver. This continues until the RV responds with a *Response* message, or a 1s negotiation timeout elapses. If no response is received within this timeout period, the HV cancels the negotiation. Otherwise, a successful negotiation transitions the HV to the execution state. In this state, the HV sends *Confirmation* messages (every 100ms) to the RV, acknowledging the RV's agreement and confirming its own execution of the maneuver. This continues until the maneuver is complete or an execution timeout occurs. The execution timeout is calculated by adding a 3s margin to the HV's initial estimate of maneuver completion time, allowing for potential deviations from the planned maneuver.

Although seamless maneuver coordination and execution are the goals, both the negotiation and execution phases are vulnerable to failure. This can lead to three distinct outcomes: successful coordination, unsuccessful negotiation, and unsuccessful execution [8]. Successful coordination means the HV completes the maneuver as intended. Unsuccessful negotiation may be due to message transmission errors or the RV's involvement in another coordination process. Unsuccessful execution arises when the maneuver exceeds the allocated time for the maneuver or the HV changes its maneuver due to unexpected traffic conditions.

## IV. Evaluation

### A. Simulation framework and parameters

We evaluate the performance of coordinated lane change maneuvers using the integrated traffic and V2X simulation platform presented in [7]. This platform incorporates the detailed maneuver coordination triggering and design presented in this paper and also described in [7]. The platform relies on ns-3 for V2X communication simulation using IEEE 802.11p and the VANET Highway Mobility module [9] for vehicle mobility modeling. The exchange of messages for intent sharing and maneuver coordination negotiation follows the *Tracking Trajectories* generation rules presented in [10]. According to these rules, *Intent* messages containing the current planned trajectory are generated and broadcast by vehicles when the trajectory has changed significantly since the last shared Intent, or if more than one second has passed since the previous message. These rules ensure that the coordination messages described in Section III.B (*Request*, *Response*, and *Confirmation*) are transmitted with high reliability, reducing channel load while maintaining intent-sharing efficiency. To avoid that maneuver coordinations affect each other, the minimum distance between coordinations is set to 500m, i.e. a maneuver coordination cannot be triggered if there is an active coordination in a range of 500m.

The simulations are conducted under two distinct highway scenarios: one involving an obstacle that may disrupt traffic flow, and another without any obstacles.. Both scenarios consist of a 5km, six-lane highway (three lanes per direction) with periodic boundary conditions. Two traffic density levels were simulated: 15 and 25 vehicles per kilometer per lane. The traffic stream comprised of 80% cars and 20% trucks. Each vehicle's desired speed (free-flow speed) is randomly selected from a uniform distribution with a ±20% variation around a mean desired speed. The mean desired speeds were 120 km/h for cars and 80 km/h for trucks. In the scenario without obstacles, vehicles can freely change lanes, except for trucks, which are restricted to the middle and right lanes. In the scenario with obstacles, a stationary vehicle is placed at kilometer 2.5 in the rightmost lane, acting as an obstacle. The stationary vehicle forces vehicles traveling in that the rightmost lane to shift to the middle lane in order to overtake it. As a result, vehicle movements—and consequently, the traffic patterns to be predicted—differ between the two scenarios. Therefore, the results will evaluate the effectiveness of the baseline prediction approach versus the intent-sharing prediction approach under the varying conditions introduced by the two scenarios and the two traffic density levels tested.

### *B. Results and discussion*

The evaluation quantifies the number of successful and unsuccessful maneuver coordinations. The number of unsuccessful negotiations was negligible due to two factors: low message transmission failure rates and the minimum distance requirement between vehicles initiating coordination requests. This distance prevented vehicles from attempting to coordinate with nearby vehicles already participating in another coordination. Consequently, unsuccessful coordinations have been produced only due to failures during the execution phase of the coordinated maneuver, mostly due to unpredicted traffic conditions.

Table I reports the number of successful and unsuccessful coordinations per vehicle and hour in the scenario without obstacles. Results show that the intent-sharing based prediction approach significantly outperforms the baseline approach and reduces the number of unsuccessful coordinations by more than two-thirds at both densities. Minimizing unsuccessful coordinations is crucial for preventing unnecessary traffic disruptions. Concurrently, the number of successful coordinations increased with the intent-sharing based prediction approach, notably at the higher density of 25 vehicles/km/lane. This increase is desirable, as prior work has demonstrated the positive impact of coordinations on traffic flow [7]. These results show that the use of intent-sharing significantly benefit maneuver coordinations and translates directly into a substantial improvement in coordination effectiveness.

Table II compares the percentage of successful and unsuccessful coordinations achieved with the baseline and the intent-sharing based prediction approaches, also for the scenario without obstacles. The results show a substantial increase in successful coordinations with intent-sharing. At the lower vehicle density, the percentage of successful coordinations increases from 57% to 85% with the intent-sharing based prediction approach, while at the higher density, it is nearly doubled from 41% to 79%. This highlights the strong potential of intent-sharing for improving maneuver coordination success rate. Despite these improvements, some unsuccessful coordinations persist with the intent-sharing based prediction approach. These failures are primarily due to the limited prediction range ($d_{pred}$, set to 500m) used in this approach. Traffic variations occurring beyond this distance, such as lane changes or decelerations, are not incorporated into the predictions. These unpredicted variations can then affect the trajectories of following vehicles, leading to unpredicted traffic variations.

The benefits achieved with intent-sharing are also shown in the accuracy of the maneuver execution. Fig. 3 shows, for all the successful coordinations in the scenario without obstacles, the PDF (Probability Density Function) of the difference between the intended Vego lane change time and the actual lane change time. As it can be observed, for most of the successful coordinations produced with intent-sharing, the intended and actual lane change time are very similar (i.e. close to zero difference). On the other hand, the difference is higher when using the baseline prediction approach. In the latter case the actual lane change time is either lower or higher than the initially intended lane change time. Our design and implementation of maneuver coordination [8] is resilient to some variation in the final lane change time. However, the trends observed in Fig. 3 shows that the use of intent-sharing further improves the accuracy of maneuver coordinations.

TABLE I. NUMBER OF SUCCESSFUL AND UNSUCCESSFUL COORDINATIONS. SCENARIO WITHOUT OBSTACLES

| Density (vehs/km/lane) | Prediction approach | # of successful coordinations (/vehicle/hour) | # of unsuccessful coordinations (/vehicle/hour) |
|---|---|---|---|
| 15 | Baseline | 4.1 | 3.1 |
| | Intent-sharing | 4.4 | 0.7 |
| 25 | Baseline | 5.1 | 5.2 |
| | Intent-sharing | 6.2 | 1.6 |

TABLE II. PERCENTAGE OF SUCCESSFUL/UNSUCCESSFUL COORDINATIONS. SCENARIO WITHOUT OBSTACLES

| Density (vehs/km/lane) | Prediction approach | % of successful coordinations | % of unsuccessful coordinations |
|---|---|---|---|
| 15 | Baseline | 57.3 | 42.7 |
| | Intent-sharing | 85.2 | 14.8 |
| 25 | Baseline | 41.0 | 59.0 |
| | Intent-sharing | 79.1 | 20.9 |

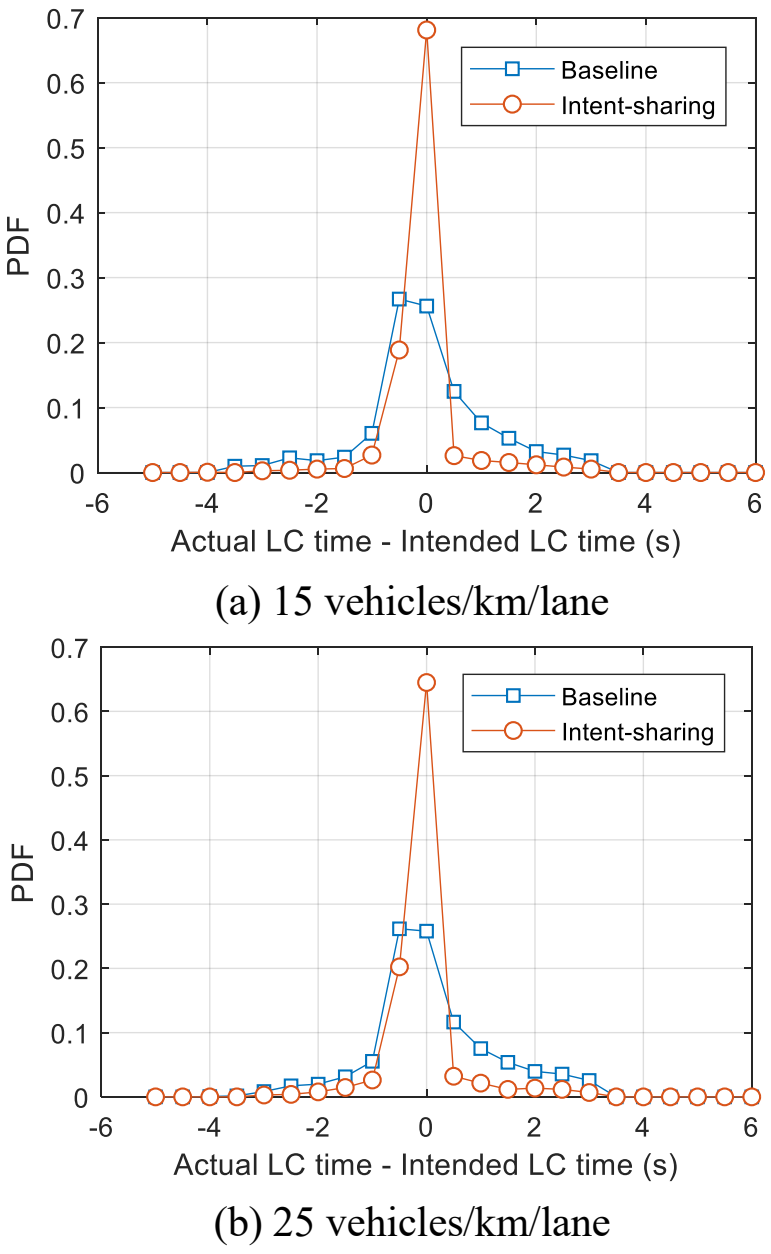

(a) 15 vehicles/km/lane

(b) 25 vehicles/km/lane

Fig. 3. PDF (Probability Density Function) of the difference between the intended and actual lane change (LC) time for the two prediction approaches and two traffic densities evaluated. Scenario without obstacles.

The results obtained in the scenario with an obstacle in the right lane exhibit similar trends to those observed in the obstacle-free scenario. Table III presents the number of successful and unsuccessful coordination attempts for this scenario. As can be observed, the total number of coordination attempts is higher for both prediction approaches at a density of 15 vehicles/km/lane, and lower at 25 vehicles/km/lane. These changes reflect the impact of the obstacle on traffic dynamics. At 15 vehicles/km/lane, vehicles are more likely to initiate lane changes to avoid the obstacle, leading to an increased number of coordination requests. In contrast, at 25 vehicles/km/lane, the obstacle causes localized congestion, making coordinated maneuvers more difficult to execute without requiring excessive deceleration. As described in Section III.A, the initiating vehicle evaluates in advance whether the coordination is feasible. Consequently, maneuvers that are not viable due to congestion are not initiated, resulting in a lower total number of coordination attempts. Nonetheless, the difference in performance between the prediction approaches remains consistent with that of the obstacle-free scenario, demonstrating that the intent-sharing approach yields performance improvements under varying traffic conditions.

TABLE III. NUMBER OF SUCCESSFUL AND UNSUCCESSFUL COORDINATIONS. SCENARIO WITH OBSTACLE IN THE RIGHT LANE

| **Density (vehs/km/lane)** | **Prediction approach** | **# of successful coordinations (/vehicle/hour)** | **# of unsuccessful coordinations (/vehicle/hour)** |
|---|---|---|---|
| 15 | Baseline | 4.4 | 3.6 |
| | Intent-sharing | 5.3 | 1.0 |
| 25 | Baseline | 4.5 | 5.0 |
| | Intent-sharing | 6.0 | 1.6 |

TABLE IV. PERCENTAGE OF SUCCESSFUL/UNSUCCESSFUL COORDINATIONS. SCENARIO WITH OBSTACLE IN THE RIGHT LANE

| **Density (vehs/km/lane)** | **Prediction approach** | **% of successful coordinations** | **% of unsuccessful coordinations** |
|---|---|---|---|
| 15 | Baseline | 55.3 | 44.7 |
| | Intent-sharing | 84.6 | 15.4 |
| 25 | Baseline | 47.3 | 52.7 |
| | Intent-sharing | 78.9 | 21.1 |

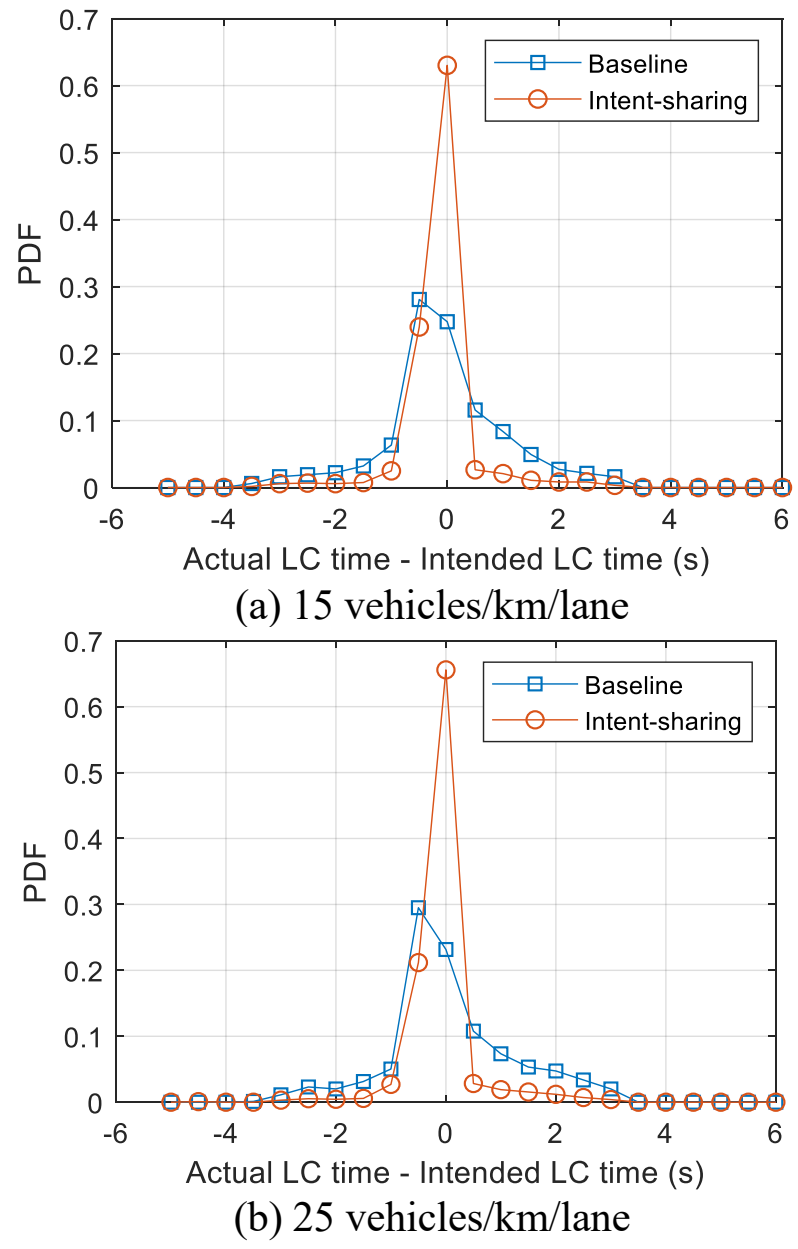

Fig. 4. PDF (Probability Density Function) of the difference between the intended and actual lane change (LC) time for the two prediction approaches and two traffic densities evaluated. Scenario with obstacle in the right lane.

Table IV presents the percentage of successful and unsuccessful coordination attempts in the scenario with the obstacle in the right lane, showing results that are consistent with those observed in the obstacle-free scenario and that demonstrate the potential of intent-sharing for a more effective maneuver coordination. Finally, Fig. 4 illustrates the PDF of the difference between the actual and the initially intended lane change times for the scenario with the obstacle. This figure highlights a significant improvement in the accuracy of the predicted lane change time in this scenario as well when the intent-sharing prediction is considered.

## V. CONCLUSIONS

Maneuver coordination relies on a vehicle's ability to assess the safety and potential benefits of the maneuver. The effectiveness of maneuver coordination is therefore strongly linked to the capacity of vehicles to accurately predict the trajectories of nearby vehicles. This study demonstrates that the direct exchange of planned trajectories through intent-sharing – instead of having to predict or infer these trajectories from current state measurements – is essential for improving the trajectory prediction accuracy, making it a critical component for effective maneuver coordination. Our evaluation shows that, in the case of coordinated lane changes in a highway scenario, intent-sharing significantly increases the number and percentage of successful coordinations. Intent sharing also improves the accuracy of maneuver execution. These improvements are observed when compared to predicting the trajectories of nearby vehicles using a kinematics-based approach instead of intent-sharing. The main motivation behind this work stems from the lack of large-scale, quantitative evaluations assessing the benefits of intent-sharing in trajectory prediction and maneuver coordination. Given that current SAE and ETSI standards consider intent-sharing optional, our findings offer valuable evidence that could inform future standardization efforts and support a more prominent role for intent-sharing in cooperative automated driving.